\documentclass[AMA,LATO1COL]{WileyNJD-v2}
 \usepackage{palatino}
\fontfamily{phv}
\selectfont
\usepackage{amssymb}

\articletype{Article Type}%

\received{ ?? April 2020}
\revised{-}
\accepted{-}

\def\reffig#1{Fig.~\ref{fig:#1}}

\def\refsec#1{\S\ref{sec:#1}}
\def\refapp#1{Appendix~\ref{app:#1}}

\begin{document}

\title{Replacing wakes with streaks in wind turbine arrays} 

\author[1]{Carlo Cossu*}
%
%

\authormark{COSSU}

\address[1]{\orgdiv{Laboratoire d'Hydrodynamique \'Energetique et Environnement Atmosph\`erique (LHEEA)}, \orgname{ CNRS - Centrale Nantes}, \orgaddress{ \country{France}}}

%

\corres{*Corresponding author: \email{carlo.cossu@ec-nantes.fr}}


\abstract[Summary]{Wind turbine wakes negatively affect downwind turbines in wind farms reducing their global efficiency. The reduction of wake-turbine interactions by actuating control on yaw angles and induction factors is an active area of research.
In this study, the capability of spanwise-periodic rows of wind turbines with tilted rotors to reduce negative wake-turbine interactions is investigated through large-eddy simulations.
It is shown that, by means of rotor tilt, it is possible to replace turbine far wakes with high-speed streaks where the streamwise velocity exceeds the freestream velocity at hub height.
Considering three aligned rows of wind turbines, it is found that the global power extracted from the wind can be increased by tilting rotors of the upwind turbine rows, similarly to what already known for the case of a single column of aligned turbines.
It is further shown that global tilt-induced power gains can be significantly increased by operating the tilted turbines at higher induction rates.
Power gains are further increased for higher ratios of rotor diameters and turbine spacings to the boundary layer  height.
All these findings are consistent with those of previous studies where streamwise streaks were artificially forced by means of spanwise-periodic rows of wall-mounted roughness elements in order to control canonical boundary layers for drag-reduction applications.}


\jnlcitation{\cname{%
\author{C. Cossu}} (\cyear{2020}), 
\ctitle{Replacing wakes with streaks in wind turbine arrays}, \cjournal{Wind En.}, \cvol{2020;00:1--12}.}

\maketitle


\fontfamily{phv}
\selectfont

\section{Introduction}\label{sec1}

In wind farms, turbines impacted by wakes generated by upwind turbines experience significant reductions in the mean available wind power and increased turbulence levels.\cite{Stevens2017,PorteAgel2019} 
A significant number of design and control strategies have been proposed to alleviate these negative effects among which great interest has been recently attracted by the approach where the rotor yaw angle is controlled in order to deflect the wake away from downwind turbines.
In yawed turbines, indeed, the misalignment of the mean thrust force and wind direction induces a pair of vertically-stacked counter-rotating vortices which increasingly deflect the wake away from the mean wind axis in the horizontal plane.\cite{Dahlberg2003,Howland2016,Bastankhah2016} 
The thrust-wind misalignment reduces the amount of power produced by the yawed turbine but  this power loss can be more than compensated by the power gain of downwind turbines induced by the wake deflection.\cite{Medici2006,Jimenez2010}

In complement to yaw control, which is associated to wake deflections in the horizontal plane, it has been recently shown that turbine wakes can be deflected in the vertical direction by acting on the rotor tilt angle\cite{Guntur2012,Fleming2014,VerHulst2015} and that the power gain in downwind turbines can be larger than the power reduction associated to the tilt of upwind turbines\cite{Fleming2015,Annoni2017,Bay2019b}.
Best performances were obtained with positive tilt angles for which the wake is deflected towards the ground. 
Furthermore, power gains obtained by means of tilt were found to be potentially larger than those associated to yaw control because the downwash associated to the positive tilt exploits the vertical wind shear by displacing downwards higher-altitude higher-velocity fluid towards downwind turbines therefore increasing their available wind power.

Despite its potential, however, tilt control has been the subject of only a few studies probably because in most of installed wind turbines, which have upwind-aligned rotors, positive tilt would lead the rotor to hit the tower. However, positive tilt capabilities are compatible with turbines with downwind-aligned rotors which are being revisited as a very promising concept because they are resilient in extreme wind situations and are compatible with highly flexible very large blades. Indeed, downwind-aligned turbines admit favourable distributions of blade bending loads and benefit from passive yaw control capabilities which are critical in off-grid situations experienced in extreme wind conditions.\cite{Kiyoki2017,Loth2017}

Previous investigations of tilt control have mostly considered the effect of tilting rotors of upwind turbines in an isolated column of aligned turbines;
for the case of two aligned turbines,\cite{Fleming2015}  the best power gains were obtained for positive tilt angles $\varphi \approx 25^o$ and, for the case of three aligned turbines \cite{Annoni2017} when both upwind turbines were tilted.
A more recent study,\cite{Bay2019b} based on a reduced wind-farm model, has considered global annual power gains (for selected wind roses) of model wind farms where tilt was applied only to peripheral turbines with fixed tilt setting (not depending on the wind direction). 
It was found that tilt could produce gains of annual power production  that were larger for 5MW wind turbines than for 13MW wind turbines. 
The best gains were obtained for tilt angles smaller than what was found for isolated-column configurations (actually, it was found that a power reduction was experienced for $\varphi \approx 25^o$).

The present study complements the few previous investigation on tilt control by further considering the effect of tilt applied to spanwise-periodic rows of wind turbines.
In this case a spanwise-periodic distribution of counter-rotating quasi-streamwise vortices is forced by the tilted turbines inducing a spanwise-periodic distribution of upwash and downwash flows.

Spanwise-periodic distributions of counter-rotating vortices, when immersed in shear flows, are known to induce quasi-streamwise streaks, i.e. streamwise-elongated spanwise-alternating high-speed and low-speed regions, which are ubiquitous in transitional and fully developed turbulent shear flows.\cite{Kline1967,Schmid2001,Hutchins2007,Cossu2017,Onder2018}
The streaks are amplified via the lift-up effect\cite{Moffatt1967,Landahl1980,Schmid2001} which is a non-modal amplification mechanism\cite{Boberg1988,Butler1992,Gustavsson1991,Trefethen1993,Cossu2009,Cossu2017}
that has been exploited as a natural control amplifier for flow-control applications.
Artificially forced streaks have indeed been used to delay transition in laminar boundary layers,\cite{Cossu2002,Fransson2006} to reduce pressure drag on idealized car models at high Reynolds numbers,\cite{Pujals2010} to reduce the turbulent friction drag in pipes\cite{Willis2010} and to suppress vortex shedding in bluff-body wakes.\cite{DelGuercio2014b,DelGuercio2014c,Marant2017}
In this context, spanwise-periodic rows of tilted wind turbines display a strong similarity with spanwise-periodic rows of roughness elements used in previous experimental studies of flow control by streaks\cite{Fransson2004,Fransson2005,Fransson2006,Hollands2009,Pujals2010b} which, similarly to the tilted turbines, produce spanwise-periodic distributions of wakes and counter-rotating  streamwise vortices.
In these flow-control studies it was found that the wakes of the roughness elements, were replaced, further downstream, by high-speed streaks, i.e. regions of streamwise velocity excess. We are interested in verifying if a similar effect can be observed in the atmospheric surface layer with forcing given by tilted turbines.

In the present study we will therefore determine if high-speed coherent streaks can be forced by spanwise-periodic rows of wind turbines with tilted rotors and if the total power of model wind farms can be increased by forcing these spanwise-periodic coherent streaks.  
The effects of changing the tilt angle and the induction factor of the tilted turbines  will be also investigated as well as that of increasing the relative size and spacing of the wind turbines with respect to the boundary layer thickness.
These effects will be explored by means of large-eddy simulations where wind turbines are modeled as actuator-disks.

The paper is organized as follows. 
The formulation of the problem at hand is introduced in  \refsec{formul} and the streaky flow forced by a single spanwise row of turbines is described in \refsec{streaks}.
The effect of forcing the streaks on the power production of three rows of wind turbines is presented in \refsec{WF} where the effect of turbine size, tilt angle and induction factor are discussed.
The main results are summarized and further discussed in \refsec{concl}.
Additional details on used numerical methods are provided  in \refapp{meth}.

\section{Problem formulation}
\label{sec:formul}

We consider the flow developing around a set of wind turbines immersed in a turbulent boundary layer. The turbines are aligned with the mean wind speed at hub height (zero yaw angle).
This complex turbulent flow is simulated by means of large-eddy simulations implemented in the Simulator for On/Offshore Wind Farm Applications (SOWFA\cite{Churchfield2012}) where the flow is modeled with the filtered Navier-Stokes equations under the usual Boussinesq approximation for the effects of density variations. 
Subgrid-scale stresses are modeled with the Smagorinsky model\cite{Smagorinsky1963} and it is assumed that near the ground the flow adheres to the Monin-Obhukov similarity theory for turbulent boundary layers above rough surfaces\cite{Monin1954} by implementing appropriate stress boundary conditions.\cite{Schumann1975} 
Slip boundary conditions are enforced at the top plane $z=H$ of the solution domain.
Additional details about the used numerical methods and the discretization parameters used in the simulations are provided in \refapp{meth}.

We will limit our analysis to the case of an isothermal flow (neutral boundary layer) driven by a constant pressure gradient neglecting the effect of Coriolis acceleration. 
The results obtained under these strong assumptions are a reasonable approximation of those that would have been obtained in a neutral atmospheric boundary layer if the wind turbines remain confined to the atmospheric surface layer.\cite{Calaf2010,Goit2015,Stevens2017}

Inflow boundary conditions for the simulations are generated by means of `precursor' simulations of the turbulent boundary layer in the absence of wind turbines\cite{Keating2004,Tabor2010,Churchfield2012} where periodic boundary conditions are enforced in the horizontal plane with $L_x$-$L_y$ streamwise-spanwise periodicity (the extension of the domain).
Once a fully developed statistically stationary regime is attained, the temporal evolution of flow variables on the inflow plane is stored and then used as inflow boundary condition for the simulations with the wind turbines.
In this way, it is possible to expose the turbines to realistic inflow turbulent wind conditions.
In this study, all the simulations with turbines make use of precursor simulations run in the same numerical domain.
Results are presented for domains of height H=1km with $L_x$ ranging from 3km to 6km and $L_y$ from 500m to 3km and spatial resolutions ranging from 14m to 21.4m (see \refapp{meth}).
The domains are not large and long enough to avoid a potential locking of boundary layer large-scale motions because of the periodic boundary conditions\cite{Fishpool2009,Munters2016} but they are well adapted to illustrate the averaged effects of rotor tilt. 

The effect of wind turbines on the flow is accounted for by means of the actuator disk model (ADM) which has been shown to correctly reproduce the main characteristics of turbines wakes except in the wake formation region.\cite{Wu2011}
In the chosen ADM approach the forces exerted by turbines blades on the fluid are averaged over the whole rotor disk. 
Following previous investigations,\cite{Calaf2010,Goit2015,Munters2017} the total force exerted by each turbine on the fluid is assumed to be
$\mathbf{F}=-C'_T \rho u_n^2 A \mathbf{e}_n/2$,
where $C'_T$ is the disk-based thrust coefficient,  $\mathbf{e}_n$ is the unit vector normal to the rotor, $u_n$ is the rotor-normal wind velocity component averaged over the rotor surface of area $A=\pi D^2/4$ and $D$ is the rotor diameter. 
The force is assumed to be uniformly distributed over the rotor surface and the effects of wake rotation are neglected.
Turbines are assumed to always operate in Region II.
The power produced by each turbine is $P=C'_P \rho u_n^3 A/2$ where $C_P'= \chi C_T'$ with the coefficient $\chi=0.9$ accounting with grid effects and the power lost by wing-tip vortices\cite{Martinez2016,Munters2017} (the results presented in this paper, where ratios of powers are examined, do not depend on the specific value chosen for $\chi$).
The optimal Betz value maximizing the power output of an isolated ideal turbine in the absence of tilt and yaw is obtained for\cite{Calaf2010,Goit2015,Munters2017} $C_T'=2$.
\begin{figure}
   \centering
   \includegraphics[width=0.95\textwidth]{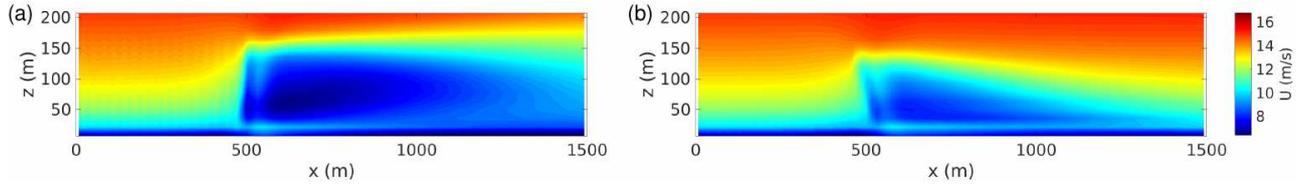}
   \vspace{-2mm}
   \caption{Mean streamwise velocity field in the vertical plane cut trough the hub axis $U(x,y_h,z)$ for the reference case 
   (panel $a$) and for the $\varphi=30^o$ tilt case (panel $b$) up to $\approx 8D$ downstream of the turbine. The flow is from left to right. The same color scale is used in both panels.
   } \label{fig:UXZvsPhiLongBox}
 \end{figure}
 
\section{Forcing streaks by tilting rotors}
\label{sec:streaks}

\subparagraph{Simulation setting}
We first consider the effect of tilt on the wakes of a (single) spanwise-periodic row of wind turbines.
The flow is simulated in a domain extending 1km in the vertical direction (which corresponds to the boundary layer thickness H) and 6km~x~0.5km  (i.e. 6H~x~H/2) in the streamwise and spanwise directions respectively.
The relatively short lateral extension of the domain enables to analyze `pure' coherent structures generated by tilted rotors excluding their interactions with boundary layer large-scale motions which have larger spanwise spacings (typically in the range $H-3H$).
The lateral domain extension corresponds to the spacing $\lambda$ of the turbines of the row.
The actuator disk dimensions (diameter $D=126m$, hub height $z_h=89m$) are based on the NREL~5-MW turbine model\cite{Jonkman2009} but, unlike that model, turbines are forced to operate in Region II for all wind speeds.
The chosen ratio $\lambda/D=4$ is equal to the one used in previous investigations of streak generation by a row of cylindrical roughness elements of diameter $D$.\cite{Fransson2004,Fransson2006,Hollands2009,Pujals2010b}
Turbines are located at $x_h=500m$ ($\approx 4D$) downstream of the inflow boundary and they operate at the constant $C_T'=1.5$, a value consistent with those observed in real wind farms.\cite{Wu2015,Stevens2015,Munters2017_Phd}
Wakes are simulated up to $\approx 40D$ downstream of the turbines.

 \begin{figure}[h]
   \centering
   \includegraphics[width=0.95\textwidth]{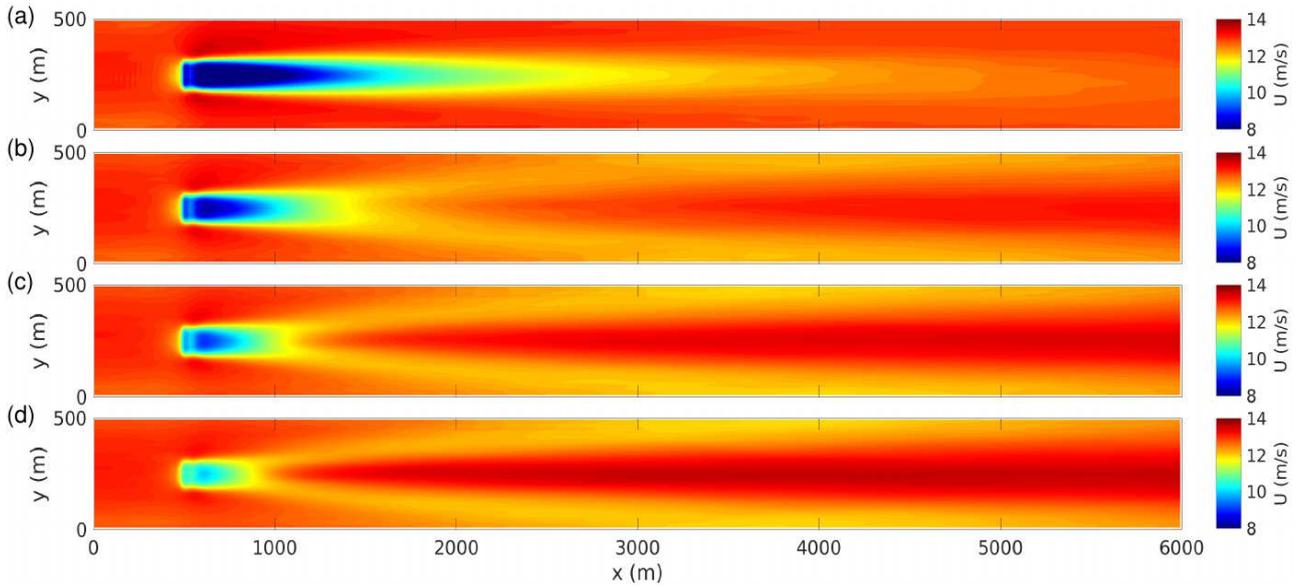}
   \vspace{-2mm}
   \caption{Mean streamwise velocity field in the horizontal plane at hub height for the reference case (panel $a$) and for the increasing tilt angles $\varphi=20^o, 30^o, 40^o$ (panels $b$ to $d$). The mean wind is from the west (the flow is left to right) and a single (spanwise) wavelength of the spanwise-periodic distribution is shown for each case. The streamwise velocity color scale is kept constant among all panels. 
   } \label{fig:UXYvsPhiLongBox}
 \end{figure}

The precursor simulation is run in the chosen domain with periodic boundary conditions allowing the complete development of the turbulent boundary layer in the absence of wind turbines with the applied constant pressure gradient $dP/dx=-0.510^{-3}~Pa/m$. 
The mean incoming wind velocity at hub height is $U_{0}=13m/s$.
The velocity and pressure fields on the inflow (west) plane are stored and used to rerun the simulation in the presence of the wind turbines which are operated starting from $t_0=20000s$. Statistics are accumulated starting from $t=24000s$ when the wakes are well developed, up to $t=40000s$.
First, the reference case is run where the rotor has the usual small negative tilt $\varphi=-5^o$ enforced to prevent any impact of the blades on the tower.\cite{Jonkman2009}
Then, the runs are repeated with unchanged parameters except for the rotor tilt angle $\varphi$.

\begin{figure}[b]
   \centering
   \includegraphics[width=0.8\textwidth]{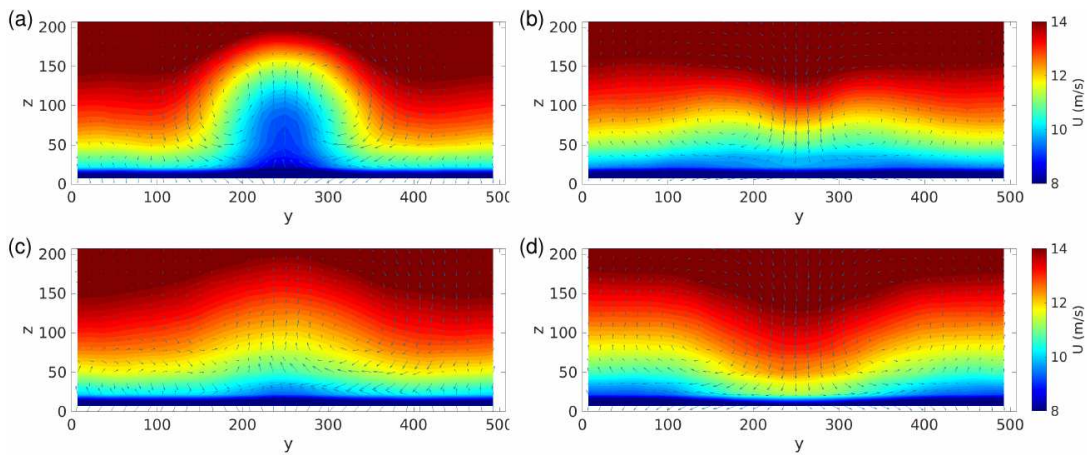}
   \vspace{-2mm}
   \caption{Mean streamwise (color scale) and cross-stream (vectors) velocity fields corresponding to the reference case (panels on the left, $a$ and $c$) and to the $\varphi=30^o$ tilted rotor case (panels on the right, $b$ and $d$) in the cross-stream planes at $x=x_h+7D$ (top row, panels $a$ and $b$) and $x=x_h+20D$ (bottom row, panels $c$ and $d$). 
   The same color scale is used in all panels for the streamwise velocity but cross-stream velocity vectors are rescaled to improved readability.
 } \label{fig:UVWYZvsPhiLongBox}
 \end{figure}
  \begin{figure}
   \centering
   \includegraphics[width=0.99\textwidth]{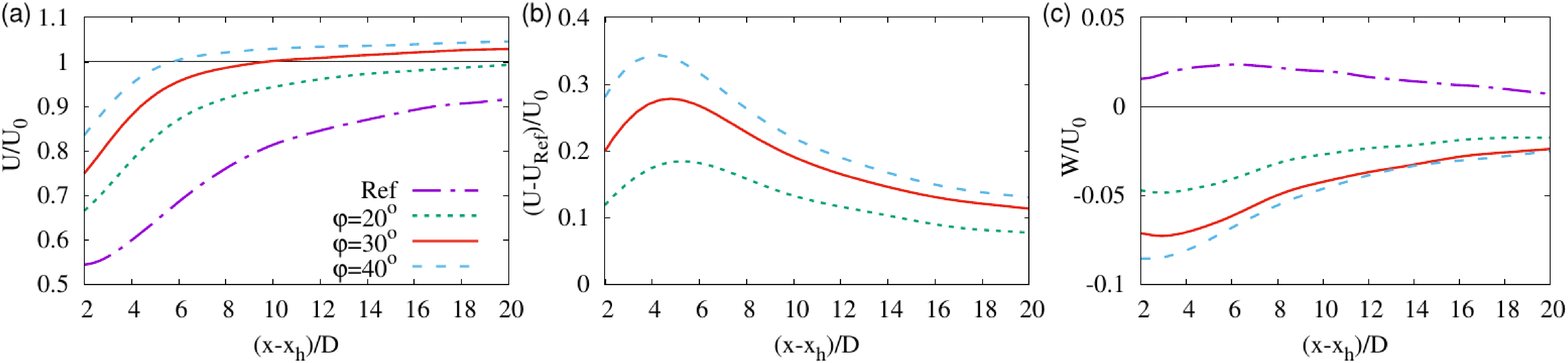}
   \vspace{-3mm}
   \caption{Downstream evolution along the hub axis of the streamwise mean velocity $U((x-x_h)/D,y_h,z_h)$ (panel $a$), the streamwise mean velocity deviation from the reference case  $[U-U_{Ref}]((x-x_h)/D,y_h,z_h)$ (panel $b$) and the vertical mean velocity $W((x-x_h)/D,y_h,z_h)$ (panel $c$) for the reference case ($\varphi=-5^o$) and for increasing rotor tilt angles $\varphi$. 
   } \label{fig:Uxofx_hub_Avg}
 \end{figure}
  
\subparagraph{Streaks formation and wake reversal} 
From \reffig{UXZvsPhiLongBox}, showing the time-averaged streamwise velocity field in the longitudinal vertical plane through the rotor axis, it can be verified that a positive rotor tilt induces the deflection of the wake towards the ground which strongly reduces its streamwise extent when compared to the reference case. 
The wake-shortening effect is also clearly visible in \reffig{UXYvsPhiLongBox} representing the time-averaged streamwise velocity field in the horizontal plane at hub height for the reference case and for increasing tilt angles.
From \reffig{UXYvsPhiLongBox} it can also be seen that for sufficiently large tilt angles the wakes are not only shortened but replaced by high-velocity regions (high-speed streaks) where the mean streamwise velocity is \emph{higher} than the mean freestream velocity at hub height ($U_0=13 m/s$).
The process by which wakes are replaced by high-speed streaks can be appreciated in  \reffig{UVWYZvsPhiLongBox} where the time-averaged velocity fields of the reference and positive tilt $\varphi=30^o$ cases are shown in the cross-stream planes situated $7D$ and $20D$ downwind of the turbine, respectively.
The two counter-rotating vortices produced by the positive rotor tilt are clearly visible just as the associated downwash which produces the wake deflection towards the ground and its replacement by the high-speed streak further downstream. 

From Figs.~\ref{fig:UXYvsPhiLongBox} and \ref{fig:UVWYZvsPhiLongBox} is can be seen that in the middle- and far-wake regions not only the low-speed region (wake) is replaced by high-speed streaks but low-speed fluid is repositioned laterally in the streamwise corridors between the turbines (recall that a single spanwise wavelength is shown in the figures).
Also, the spanwise size of the wakes and of the high-speed streaks, which are of the order of the rotor diameter $D$ in the near wake increase towards $\approx \lambda/2$ (half the turbine separation) in the far wake, as can also be appreciated from \reffig{UXYvsPhiLongBox}. 
These features are similar to those observed when streaks are forced by roughness elements in canonical laminar and turbulent flat-plate boundary layers\cite{White2002,Fransson2004,Pujals2010b} where high-speed streaks emerge downwind of the roughness elements and low-speed streaks replace the high-speed regions in the corridors.

From \reffig{Uxofx_hub_Avg}$(c)$ it can be seen that the downwash associated to the 
counter-rotating vortices generated by the tilted rotors decays downstream, just as observed in flat-plate boundary layers. 
However, in the present case the rotor tilt has two separate important effects on the wake in what concerns streamwise velocities.
The first, direct, effect of tilt is the reduction of the initial velocity deficit in the wake associated to the reduction of the streamwise component of the thrust vector, which induces reductions of the extracted wind power.
The second, indirect, effect is that the forced counter-rotating vortices redistribute momentum in the vertical direction increasing speeds downstream of the rotor where higher-located higher-speed fluid is displaced downwards (this is the famous lift-up effect\cite{Schmid2001}). 
It is this latter effect that makes possible the replacement of the wake by high-speed streaks where the streamwise velocity is higher than the incoming mean flow speed at hub height.

Fig.~\ref{fig:Uxofx_hub_Avg}$(b)$ also reveals that the tilt-induced maximum absolute gain in streamwise velocity with respect to the reference case is obtained roughly four diameters, i.e. one spanwise wavelength $\lambda=4D$ downstream of the turbine and is relevant up to $\approx 8D = 2\lambda$ downstream. 
The downstream region where best power gains could be obtained ($x-x_h\approx 4D-8D$) is therefore situated upstream of that where U is highest ($x>20D$) or even that where U exceeds the freestream velocity ($x>9D$ for $\varphi=30^o$) because power gains depend on $ U-U_{Ref}$ and not simply on $U$.

\section{Forcing high-speed streaks to increase global efficiency}
\label{sec:WF}

\begin{figure}
   \centering
   \includegraphics[width=0.95\textwidth]{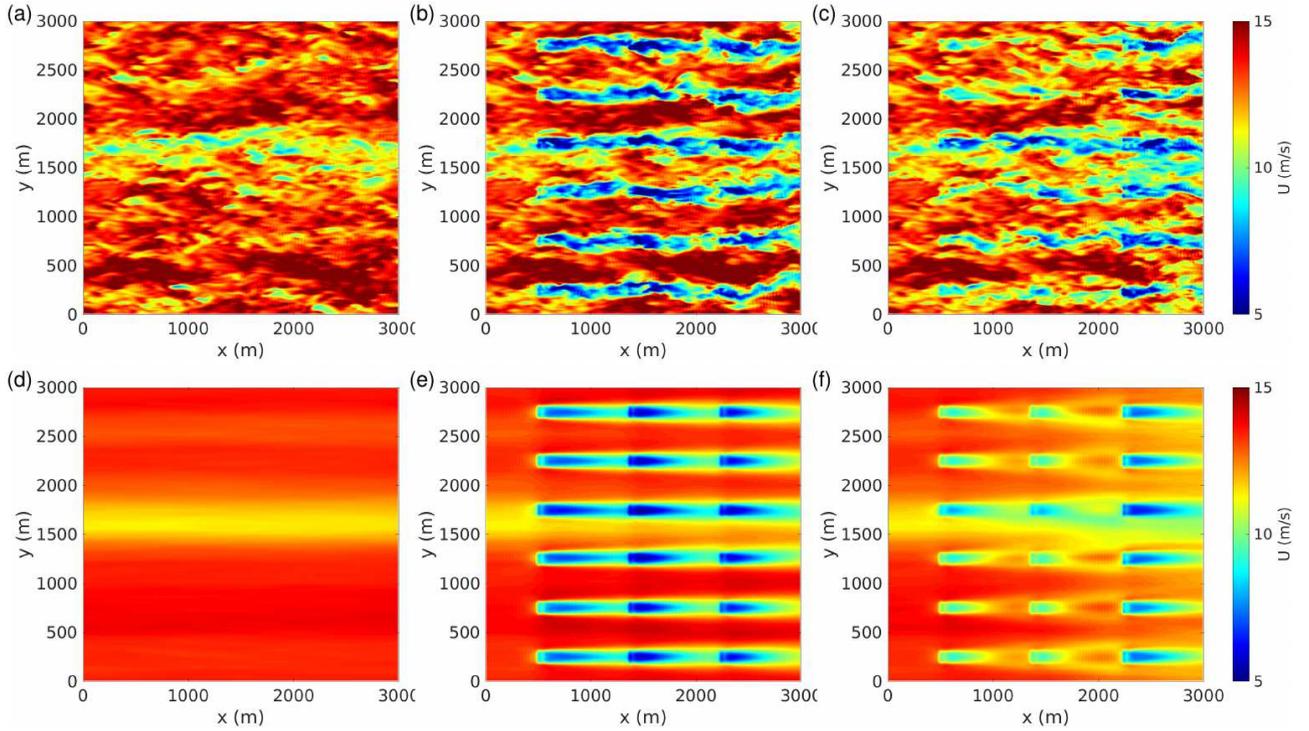}
   \vspace{-2mm}
   \caption{Instantaneous (top panels $a$, $b$, $c$) and time-averaged (bottom panels $d$, $e$, $f$) streamwise velocity field in the horizontal plane at hub height $z_h=89m$ for the precursor simulation (left panels $a$, $d$), the reference case (middle panels $b$, $e$) and the case where rotors of the upwind and middle rows are tilted by $\varphi=30^o$ (right panels $c$, $f$). The same color scale is used in all panels. All turbines have $D=126m$ and are operated at $C'_T=1.5$. The flow is from the west (left to right). The signature of a persistent large-scale boundary layer low-speed streaks is clearly discernible near $y \approx 1600$ in all panels.} \label{fig:UXZvsPhi5MW}
 \end{figure}
\begin{figure}[b]
   \centering
   \includegraphics[width=0.65\textwidth]{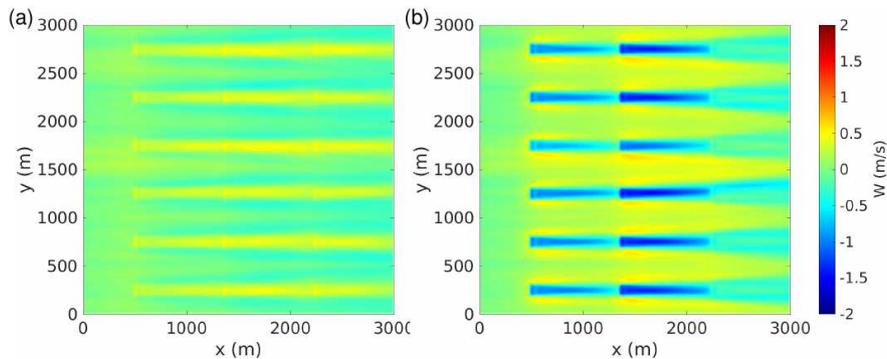}
   \vspace{-2mm}
   \caption{Time-averaged vertical velocity field in the horizontal plane at hub height $z_h$ for the reference case (panel $a$) and the case where rotors of the upwind and middle rows are tilted by $\varphi=30^o$ (panel $b$).  The same color scale is used all panels. Turbines have $D=126m$ and are operated at $C'_T=1.5$.} \label{fig:WXZvsPhi5MW}
 \end{figure}

\subsection{Influence of tilt angles and thrust coefficients on power gains}
\label{sec:WF1}

\subparagraph{Simulation setting}
We now consider the potential power gains that can be obtained by tilting rotors of upwind turbines in multiple-rows configurations.
Preliminary tests show that, similarly to the case of two turbines,\cite{Fleming2015} moderate global power gains 
can be obtained with two rows of wind turbines by tilting the rotors in the upwind row.
A more recent study has shown that, in the case of three turbines aligned in a single column, higher power gains 
could be obtained when both the upwind and the middle rotors are tilted.\cite{Annoni2017}
New simulations are therefore performed with three rows of $D=126m$ turbines with $4D$ spanwise spacing in each row and wind-aligned corresponding turbines.
Turbines rows are spaced by $7D$ in the streamwise direction (as in previous studies of tilt control of a single column of two and three turbines\cite{Fleming2015,Annoni2017}).
This value is large enough to attain sufficient absolute values of the mean streamwise velocity in the wake of upwind turbines (see \reffig{Uxofx_hub_Avg}$a$) but it does not exceed by a too large amount the value ($\approx 4-6D$) where the extra velocity recovery due to the tilt is maximum (see \reffig{Uxofx_hub_Avg}$b$).

Simulations are run in a domain with the same height (1km) and pressure gradient as in \refsec{streaks} but with a different 3km x 3km horizontal extension which enables the development of large-scale motions in the turbulent boundary layer\cite{Hutchins2007,Cossu2017} allowing for reliable statistics of turbines power production.
A precursor simulation is run in the same domain and used to generate the inflow boundary conditions that are used for the simulation in the presence of the turbines.
The presence of a large-scale coherent boundary layer low-speed streak can indeed be clearly discerned in \reffig{UXZvsPhi5MW}. 

Simulations are then performed in the presence of the 3 aligned rows of 6 turbines which can be accommodated in the simulation domain  (see \reffig{UXZvsPhi5MW}) with the chosen streamwise and spanwise spacing (but an infinite number of turbines is effectively considered in the spanwise direction because of the spanwise-periodic boundary conditions).  
The upwind row is situated $4D$ downstream of the inflow boundary where the mean incoming wind velocity at hub height is $U_0 \approx 13m/s$.
Statistics are accumulated from t=24000s (more than one hour after the turbines are switched on at t=20000s) to t=30000s.

First, the reference case is simulated with all turbines operating in the same conditions ($\varphi=-5^o$, $C'_T=1.5$). 
For this case, the usual situation where the wake of upwind turbines strongly reduces the mean wind seen by the aligned downwind turbines is observed, as shown in Figs.\ref{fig:UXZvsPhi5MW}$(b)$ and \ref{fig:UXZvsPhi5MW}$(e)$.
Preliminary computations (not shown) indicate that the best power gains are obtained when the rotors of both the upwind and the middle rows of turbines are tilted. 
The runs are therefore repeated with all parameters unchanged except for the rotor tilt of the upwind and middle rows turbines.
Results are shown for the case where they are tilted by the same angle $\varphi$, for selected values of $\varphi$.

\subparagraph{Influence of tilt control angle at fixed thrust coefficient}
Instantaneous and mean streamwise velocity fields  are reported in \reffig{UXZvsPhi5MW} in the plane at hub height for the precursor simulation, the reference case and the case with $30^o$ tilt of the rotors of the upwind and middle turbine rows.
Mean vertical velocity fields at hub height are reported in \reffig{WXZvsPhi5MW}.
From these figures it can be seen that, as already observed in the case of three aligned turbines\cite{Annoni2017} and two aligned rows of roughness elements in a plane channel,\cite{Hollands2009} the effects of rotor tilt in the two upwind rows are almost additive resulting in vertical downwards velocities (downwash) and wake recoveries which are stronger downwind of the middle row of turbines than downwind of the front (most upwind) row.

 \begin{figure}
   \centering
   \includegraphics[width=0.75\textwidth]{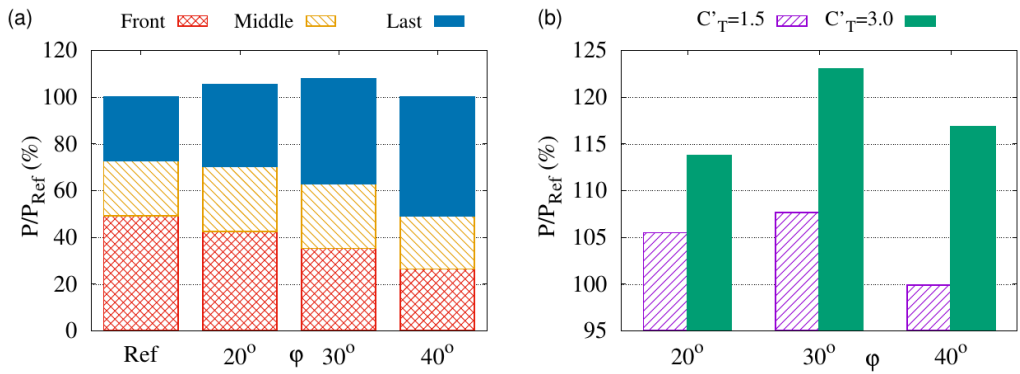}
   \vspace{-2mm}
   \caption{Influence of the tilt angle $\varphi$ on the average power extracted by the 6x3 turbine array.
   Panel $(a)$: percentage ratio of the produced power $P$ to the $P_{Ref}$ power produced for the reference case decomposed into the contributions of the upwind tilted (cross-hashed, red), middle tilted (hashed, yellow) and downwind not tilted (solid, blue) turbine rows when all turbines are operated at $C'_T=1.5$. 
   Panel $(b)$: comparison of the $P/P_{Ref}$ dependence on $\varphi$ for the $C'_T=1.5$ case (same case as in panel $a$, hashed, magenta) and for the case where the tilted turbines are operated at $C'_T=3$ (solid, green).
 } \label{fig:PvsPhi}
 \end{figure}
  
In \reffig{PvsPhi}$(a)$ the mean total power $P$ produced by the three rows of turbines when rotors of the two upwind rows are tilted is compared to the mean power $P_{Ref}$ produced in the reference condition. From this figure it is seen that the effect of increasing the positive rotor tilt $\varphi$ is to decrease the mean power extracted by the most upwind row of turbines (because of the reduction of the normal momentum flux trough the tilted rotor) and to increase that of the last (most downwind) row of turbines (because of the increase of the mean streamwise velocity on the rotor).
A milder variation is observed for the power extracted by the middle row where the two contrasting effects are at play.
Overall, the beneficial effects overcome the detrimental effect of the tilt resulting in a  
total power increase with respect to the reference case, which is maximal for $\varphi \approx 25^o-30^o$.

\subparagraph{Influence of an increased induction in tilted turbines}
We note that an increase of $C_T'$ from 1.5 to 3 in the tilted turbines results in an increased vertical component of the thrust enhancing the streamwise vortices.
The simulations with tilted rotors have been therefore  repeated by operating the tilted rotors at $C_T'=3$ while leaving the turbines with non-tilted rotor at the nominal $C'_T=1.5$.
The results, reported in \reffig{PvsPhi}$(b)$, show that operation of the tilted rotors at $C'_T=3$ leads to a substantial increase of the tilt-induced power gain, with an almost tripled maximum power gain obtained near $\varphi=30^o$ .
This indicates that there certainly is room for further enhancement of tilt-induced power gains by optimizing $\varphi-C'_T$ combinations in wind farm operation. We leave such an optimization for future study.

\subsection{Influence of the relative rotor size on power gains}

 \begin{figure}[b]
   \centering
   \includegraphics[width=0.95\textwidth]{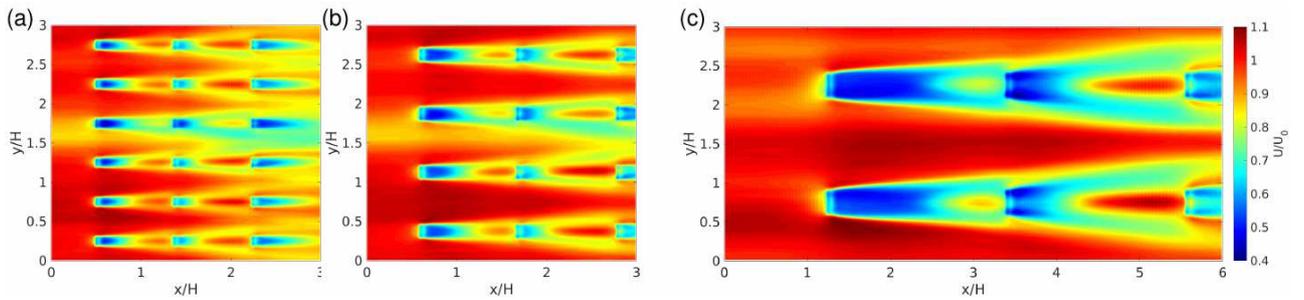}
   \vspace{-2mm}
   \caption{Effect of turbine size relative to the boundary layer thickness on the streamwise velocity in the horizontal plane at hub height for $(a)$~the previously considered $D=126m$ turbines in the 1km-thick boundary layer ($D/H=0.126$), $(b)$~the intermediate ratio $D/H=0.18$, and $(c)$~the largest ratio $D/H=0.36$. 
   Upwind and middle rows turbines are operated with $\varphi=30^o$ and $C'_T=3$. 
   Turbines of the most downwind (eastward) row are operated at reference parameter values. The mean wind is from the west (left to right). Same color scale in all panels.
   } \label{fig:UXY_AllTurbs}
 \end{figure}

The computation of optimal perturbations of canonical turbulent wall-bounded flows indicates that the largest energy amplifications of coherent streamwise streaks can be attained when the spanwise  spacing of the streaks and of the vortices used to force them is of a few boundary layer thicknesses,\cite{Cossu2009,Pujals2009,Hwang2010c} requiring roughness elements with diameters of the order of the boundary layer thickness.\cite{Pujals2010b,Pujals2010}
The results reported in the previous sections, obtained with actuator disks based on the NREL 5-MW turbine dimensions, correspond to a ratio $D/H=0.126$ which is an order of magnitude smaller than  optimal ratios. 
In fact, optimal $D/H=O(1)$ ratios can not be considered in the present setting  where the boundary layer thickness coincides with the vertical extension of the solution domain where (horizontal) slip boundary conditions are enforced.
However, even if relatively far from the values of optimal spanwise spacing, a moderate increase of the amplification of the streaks issued from quasi-streamwise vortices of given energy can be expected if the $D/H$ ratio is, even moderately, increased.

We therefore consider two reasonably larger ratios $D/H=0.18$ (approximately corresponding to the DTU 10-MW turbine model\cite{Bak2013} with $D=178m$ immersed in a 1km-thick boundary layer) and $D/H=0.36$ for three rows of turbines keeping constant to $4D$ and $7D$ their relative spanwise and streamwise spacing. 
Additional simulations are therefore performed with 4x3 turbines with $D/H=0.180$ in the 3km x 3km x 1km domain considered in \refsec{WF1} and with 2x3 turbines with $D=360m$ in an additional 6km x 3km x 1km domain as shown in \reffig{UXY_AllTurbs}.
Building on the previous findings, the same tilt angle is enforced on all rotors of the  upwind and middle row turbines which are operated with $C'_T=3$, while the most downwind row is operated at reference values.

\begin{figure}[h]
   \centering
   \includegraphics[width=0.38\textwidth]{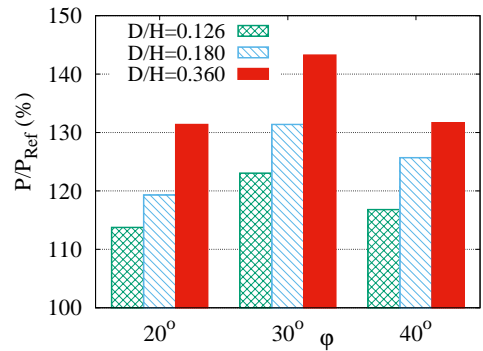} 
   \vspace{-3mm}
   \caption{Influence of the turbine diameter to boundary layer height ratio $D/H$ on power gains  $P/P_{ref}$ when tilted-rotor turbines are operated at $C'_T=3$.
}\label{fig:PvsPhiDH}
\end{figure}

The additive wake-shortening effect induced by rotor tilts is seen to operate similarly for all the considered $D/H$ ratios, as the relative spanwise and streamwise turbine spacing have been kept constant.
As anticipated, mean power gains are increased for larger $D/H$ ratios, as reported in \reffig{PvsPhiDH}, up to a maximum observed power gain exceeding 40\% for the largest considered value of $D/H$.
Despite their very encouraging nature, however, the latter results should be taken with care because for $D/H \gtrsim 0.2$, turbines are no more confined into the logarithmic region of the considered pressure-driven boundary layer which has a structure similar to that of the atmospheric surface layer. In this case, streak amplifications cease to be `universal' but depend on the particular structure of the flow in the outer layer.
Additional work is therefore needed to confirm and extend the results obtained for the largest $D/H$ ratio in more realistic atmospheric boundary layer settings.  

\vspace{-3mm}

\section{Conclusion}
\label{sec:concl}

This main goals of this study were to determine if rotor tilt in spanwise-periodic rows of wind turbines could be used to replace the (low speed) wakes of the turbines with high-speed streaks and determine the global power gains that could be obtained by forcing these streaks in order to enhance the power production of downwind turbines. 

It is found that spanwise-periodic rows of wind turbines with tilted rotors can be effectively used to force spanwise-periodic distributions of quasi-streamwise vortices which, for sufficiently large positive tilt angles, are able to  `reverse' the wakes by  replacing them with high-speed streaks.
The observed wake-reversal is similar to that documented in previous investigations where  spanwise-periodic rows of cylindrical roughness elements were used to force streaks in canonical flat-plate boundary layers.\cite{White2002,Fransson2004,Fransson2006,Pujals2010b,Pujals2010}
However, while streamwise velocities higher than the mean incoming freestream velocity at hub height can be found in the far wake, the maximum of the mean streamwise velocity recovery induced by the rotor tilt is attained more upstream ($\approx 4-5D$ downwind of the rotor). 

The analysis of the power production of three spanwise-periodic rows of aligned wind turbines confirms that the total mean power can be increased by tilting the rotors in the front (upwind) and the middle turbine rows, with best performances obtained when $\varphi \approx 25^o-30^o$ for the considered configurations. 
The relative power gains obtained with actuator disks with same dimensions as the NREL~5-MW turbines and operated at $C'_T=1.5$ 
are similar to those found for an isolated column of three NREL~5MW turbines\cite{Annoni2017} by including, unlike in this study, the effects of wake rotation, radial distribution of the aerodynamics forces on the blades, Coriolis acceleration and the capping inversion. This demonstrates the robustness of the mechanisms underlying power gains obtained by tilting rotors.

It was then verified if further power gains could be obtained by increasing the thrust coefficient of the tilted turbines. 
It is found that operating the tilted turbines at $C'_T=3$ 
instead of $C'_T=1.5$, tilt-induced power gains can be highly increased.
It is believed that additional power gain improvements can be achieved by means of a systematic optimization of $\varphi-C'_T$ distributions for tilted turbines that we leave for future study.

It is additionally shown that further substantial tilt-induced power gains are obtained with ratios $D/H$ and $\lambda/H$ of turbines diameter and spanwise-spacing to the boundary-layer height larger than those of the NREL~5MW turbines immersed in a 1km-thick boundary layer ($D/H=0.126$, $\lambda/H=0.5$).
This is consistent with previous results showing that maximum amplifications of coherent large-scale streaks in turbulent boundary layers are obtained for $\lambda/H$ ratios of a few units.\cite{Pujals2009,Cossu2009,Hwang2010,Willis2010}
Further investigations are, however, needed to confirm and extend these results to the $D/H=O(1)$ regime where realistic atmospheric boundary layer profiles and the effect of Coriolis acceleration must be taken into due account.
This high-$D/H$ regime is not only of interest for futuristic very-large turbines but also for current-generation turbines operating in shallow atmospheric boundary layers (for instance, the Haliade-X diameter $D=220m$ is comparable to typical nocturnal boundary layer heights).  
In the case of (shallow) stable boundary layers, furthermore, the higher vertical velocity gradients, promoting more efficient streak amplification, coupled with reduced turbulent levels in the incoming flow, associated to poorer wake recovery in the reference case, might lead to dramatic power gains. This is the subject of a current intense research effort.

Additional investigations are also needed to explore the benefits of tilt control in deep turbine arrays of large wind farms and for a complete range of wind directions.
Configurations with peripheral tilted turbines acting on a wind-farm have been very  recently investigated\cite{Bay2019b} for constant-tilt zero-yaw and constant $C_P$ operation mode. 
It would be very interesting to extend such type of investigations using optimized $\varphi-C_T$ combinations possibly complemented with yaw control to optimally target the forced high-speed streaks to downstream turbines.
Also, as the high-speed fluid in the corridors between wind turbines enhances wake recovery in the absence of tilt, it remains to be understood if its replacement with lower speed fluid induced by rotor tilt can significantly affect tilt-control performances, especially in deep turbine arrays.

\vspace{-3mm}
\section*{Acknowledgments}
\vspace{-3mm}
I gratefully acknowledge the use of the Simulator for On/Offshore Wind Farm Applications (SOWFA) developed at NREL\cite{Churchfield2012} based on the OpenFOAM finite volume framework.\cite{OpenFOAM}


\newcommand{\noopsort}[1]{} \newcommand{\printfirst}[2]{#1}
  \newcommand{\singleletter}[1]{#1} \newcommand{\switchargs}[2]{#2#1}

\appendix
\section{Methods}
\label{app:meth}

The standard numerical schemes and parameters implemented in SOWFA\cite{Churchfield2012} and built on standard OpenFOAM (release 2.4.x) solvers are used to solve the filtered Navier-Stokes equations with Boussinesq fluid model and Smagorinsky\cite{Smagorinsky1963} modeling for the subgrid scale motions. 
The PIMPLE scheme is used for time advancement. 
Schumann's stress boundary conditions\cite{Schumann1975} are enforced at the near-ground horizontal boundary.

Numerical simulations of the considered turbulent boundary layers have been performed in $L_x$~x~$L_y$~x~$H$ numerical domains of (vertical) height $H$, (streamwise) length $L_x$ and width $L_y$.
Three domains have been considered. 
Domains D1 (6km~x~0.5km~x~1km) and D2 (3km~x~3km~x~1km), both discretized with 15m~x~14m~x~14m cells, have been used for the simulations implying actuator disks with the dimensions of the NREL~5-MW ($D/H=0.126$) and DTU~10-MW ($D/H=0.18$) turbines.
Domain D3 (6km~x~3km~x~1km), discretized with 21.4m~x~20~x~14m cells, has been used for the simulations implying actuator disks with $D/H=0.36$.
The solutions are advanced with $\Delta t=0.8s$ time steps satisfying the CFL~$<$0.45 constraint and keeping reasonable the amount of data stored in precursor simulations.

The original actuator disk (ADM) turbine model implemented in SOWFA, which includes wake rotation effects as well as the blade-derived radial dependence of the forces acting on the fluid\cite{Martinez2013}, has been modified to implement the in-house ADMC model used in the present study by: 
(a) keeping the same discretization points on the disk but distributing the body force uniformly in the radial direction, 
(b) setting the body force magnitude to\cite{Calaf2010,Goit2015,Munters2017}  $\mathbf{F}=-C'_T \rho u_n^2 A \mathbf{e}_n/2$ removing its dependence on the turbine controller, 
(c) removing body force components parallel to the rotor plane (inducing wake rotation). 
In this way, the turbine response only depends on $C'_T$ and not on the turbine controller settings, simplifying the interpretation of the results.
The Gaussian projection of the control-point-discretized body forces with a smoothing parameter $\varepsilon$ is left unchanged with $\varepsilon=20m$ for simulations in domains D1 and D2 and $\varepsilon=30m$ for simulations performed in domain D3.
\end{document}